\documentclass[a4paper]{jpconf}
\usepackage{graphicx}
\begin{document}
\title{Theory and Simulation of Magnetic Materials: Physics at Phase Frontiers}

\author{H  T  Diep$^1$, Virgile Bocchetti$^1$, Danh-Tai Hoang$^2$ and V T Ngo$^3$}

\address{$^1$ Laboratoire de Physique Th\'eorique et Mod\'elisation, Universit\'e de Cergy-Pontoise, CNRS, UMR 8089, 2 Avenue Adolphe Chauvin, F-95302 Cergy-Pontoise Cedex, France\\
$^2$  Asia Pacific Center for Theoretical Physics,
POSTECH, San 31, Hyoja-dong, Nam-gu,
Pohang, Gyeongbuk 790-784, Korea\\
$^3$ Institute of Physics, Vietnam Academy of Science and Technology,
10 Daotan, Thule, Badinh, Hanoi, Vietnam}

\ead{diep@u-cergy.fr, virgile.bocchetti@u-cergy.fr, danh-tai.hoang@apctp.org, nvthanh@iop.vast.ac.vn}

\begin{abstract}
The combination of theory and simulation is necessary in the investigation of properties of complex systems where each method alone cannot do the task properly.  Theory needs simulation to test ideas and to check approximations. Simulation needs theory for modeling and for understanding results coming out from computers.  In this review, we  give recent examples to illustrate this necessary combination in a few domains of interest such as frustrated spin systems, surface magnetism, spin transport and melting. Frustrated spin systems have been intensively studied for more than 30 years.  Surface effects in magnetic materials have been widely investigated also in the last three decades. These fields are closely related to each other and their spectacular development is due to numerous applications.   We confine ourselves to theoretical developments and numerical simulations on these subjects with emphasis on spectacular effects occurring at frontiers of different phases.
\end{abstract}

\section{Introduction}
The physics at frontiers of different phases of the same system is a very exciting subject.  In general, when  there exist between the particles of a system  several interactions, each of which favors a different symmetry, the system chooses the symmetry where its internal energy is minimum if the temperature $T=0$ or where its free energy is minimum if $T\neq 0$.  Let us take a simple example:  we consider a chain of Ising spins interacting with each other via an interaction $J_1$  between nearest neighbors (NN) and an interaction $J_2$ between next NN (NNN).  If $J_1$ is ferromagnetic ($J_1>0$) and $J_2=0$ then the ground state (GS) is ferromagnetic. Now, if $J_2$ is  antiferromagnetic ($J_2<0$), we cannot arrange the spins in an order that fully satisfies at the same time $J_1$ and $J_2$: if $|J_2| \ll (\gg) J_1$ then the system chooses to satisfy $J_1$ ($J_2$) by taking the ferromagnetic (antiferromagnetic) order.  There exists a critical value $\eta_c$ of $J_2/J_1$ where the system changes from the ferromagnetic symmetry to the antiferromagnetic one.  The critical value $\eta_c$ is the frontier between the two phases.  In general, when there are more than two interactions, the determination of the frontiers between different phases is more complicated.  As we see, the competition between ``incompatible" interactions creates frontiers. This is not limited to physics. Near a frontier, physical behaviors are different on the two sides:  due to their different symmetries, they have different laws that govern fluctuations etc.  When we introduce into one phase an external perturbation such as the temperature or an applied field,   the system can choose the symmetry of the other phase. Such a phenomenon occurs very often in many systems: it is called ``reentrance".

In this review, we give various examples of physical phenomena which occur around the phase frontiers. These examples are taken from our recent and current works.

The first subject concerns frustrated spin systems where competing interactions cause many spectacular effects (see reviews given in Ref. \cite{DiepFSS2013}). This is shown in section \ref{FSS}.  We will describe there the reentrance, the disorder lines and the partial disorder which have been found in exactly solved models \cite{Diep1987,Diep1991a,Diep1991b,Diep1992}. We will show that simulations give complementary results that exact methods cannot reach.
The second subject concerns theories and simulations of magnetic thin films. This is shown in section \ref{SM}.  One of the most important surface effects is the existence of localized spin-wave modes near a surface and interface.   It has been shown a long time ago \cite{Diep1979} that low-lying surface modes affect strongly macroscopic properties of magnetic systems giving rise for instance to a low surface magnetization or surface magnetically dead layer and to surface phase transitions at low temperatures. We discuss in particular  surface effects in frustrated materials and in films with dipolar interaction \cite{Ngo2007,Hoang2013}. Recent results of  spin resistivity are shown and discussed in relation with surface disordering \cite{Akabli2008,Magnin2011}. Finally,  results on melting and surface lattice relaxation \cite{Bocchetti2013b} are also shown.
Concluding remarks are given in section \ref{Concl}.

\section{Frustrated spin systems near phase frontiers}\label{FSS}

\begin{figure}
\begin{center}
\includegraphics[width=2.0 in]{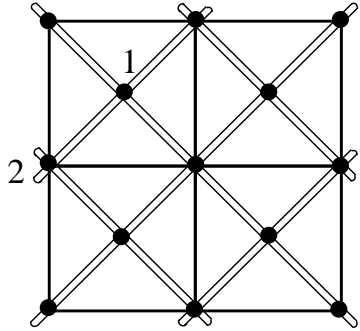}
\end{center}
\caption{\label{re-fig17}Centered square lattice. Interactions
between NN and NNN, $J_{1}$ and $J_{2}$, are denoted by white and
black bonds, respectively. The two sublattices are numbered 1 and
2.}
\end{figure}

A system is said frustrated when all interaction bonds cannot be simultaneously satisfied in the GS. Well-known examples include the triangular lattice with NN antiferromagnetic interaction and the example given in the Introduction. For definiteness, let us take the case of the ``centered square lattice" with Ising spins shown in Fig. \ref{re-fig17}, introduced by Vaks {\it et al.} \cite{Vaks} with NN and
NNN interactions, $J_{1}$ and $J_{2}$, respectively.   The exact expression for the free energy,  some
correlation functions, and the magnetization of one sublattice
were given in the original work of Vaks {\it et al.}

\begin{figure}
\begin{center}
\includegraphics[width=2.2 in]{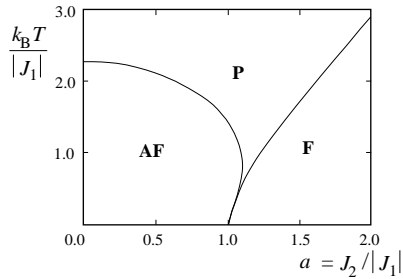}
\end{center}
\caption{\label{re-fig18}Phase diagram of centered square lattice near the ``phase frontier" $a=-1$.}
\cite{Vaks}
\end{figure}

The GS
properties of this model are as follows : for $a = J_{2}/\mid
J_{1}\mid > - 1$, spins of sublattice 2 orders ferromagnetically
and the spins of sublattice 1 are parallel (antiparallel) to the
spins of sublattice 2 if $J_{1} > 0$ ( $< 0$ ); for $a < -1$,
spins of  sublattice 2 orders antiferromagnetically, leaving the
centered spins free to flip.
The phase diagram of this model is given by Vaks  {\it et al.} \cite {Vaks}.
Except at the ``frontier" $a = -1$, there is always a finite critical temperature.
When $J_{2}$ is antiferromagnetic ($>0$) and $J_{2}/J_{1}$ is in a
small region near 1, namely near the frontier of the two phases,  the system is  successively  in the
paramagnetic state, an  ordered state, the {\bf reentrant
}paramagnetic state, and another ordered state, with decreasing
temperature (see Fig. \ref{re-fig18}).
\begin{figure}
\begin{center}
\includegraphics[width=2.2 in]{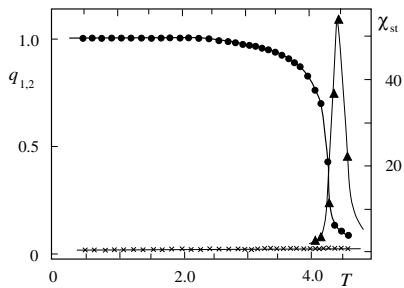}
\end{center}
\caption{\label{re-fig19} Temperature dependence of sublattice
Edwards-Anderson order parameters, $q_{1}$ and $q_{2}$ (crosses
and black circles, respectively) in the case $a = J_{2}/\mid
J_{1}\mid = - 2$, by Monte Carlo simulation. Susceptibility
calculated by fluctuations of magnetization of sublattice 2 is
also shown. The lattice used contains $N = 2\times 60 \times 60$
spins with periodic boundary conditions.}\cite{Aza89a}
\end{figure}

Though an exact critical line was
obtained \cite {Vaks}, the ordering in the
antiferromagnetic (frustrated) region has not been exactly
calculated. We have
studied this aspect by means of Monte Carlo (MC)
simulations \cite{Aza89a} which show the coexistence between order and
disorder. This behavior has been observed in three-dimensional (3D)
Ising spin models \cite{Blan,Diep85b} and in an exactly soluble
model (the Kagom\'{e} lattice) \cite{Diep1987} as well as in frustrated 3D quantum spin 
systems \cite{Quartu1997,Santa1997}. The results for the Edwards-Anderson sublattice order
parameters $q_{i}$ and the staggered susceptibility of sublattice
2 , as functions of $T$, are shown in Fig. \ref{re-fig19} in the
case $a = - 2$.

As is seen, sublattice 2 is ordered up to the transition at
$T_{c}$ while sublattice 1 stays disordered at  all $T$. This
result shows a new example where order and disorder  coexists in
an equilibrium state.  It í noted that in the paramagnetic region,  a Stephenson disorder line \cite{Ste} has
been found in Ref. \cite{Aza89a}
\begin{equation}
\cosh (4J_{1}/k_{B}T_{D}) = \exp(-4J_{2}/k_{B}T_{D})\label{eq44}
\end{equation}
The two-point correlation function at $T_{D}$ between spins of
sublattice 2  separated by a distance r is zero for odd distance $r$ and decay
like  $r^{- 1/2}[\tanh (J_{2}/k_{B}T_{D})]^{r}$  for  even $r$ \cite{Ste}.
However,there
is {\it no dimensional reduction} on the Stephenson line given above.
Usually, one defines the disorder point as the temperature where
there is an effective reduction of dimensionality so
that physical quantities become simplified spectacularly \cite{Mail}.
In general, these two types of disorder line are equivalent, as
for example , in the case of the Kagom\'{e} lattice Ising model (see below).
This is not the case here. The  disorder
line corresponding to dimensional reduction, was given for the
general 8-vertex model by Ref. \cite {Giaco86}.
When this result is applied to
the centered square lattice, one finds that the disorder variety is given by
\begin{equation}
\exp(4J_{2}/k_{B}T) = (1-i\sinh (4J_{1}/k_{B}T))^{-1}\label{eq45}
\end{equation}
where $i^{2} = -1$. This disorder line lies on the unphysical
(complex) region of the parameter space of this system. Only the
Stephenson disorder line Eq. (\ref{eq44}) is the relevant one for
the reentrance phenomenon.
Disorder solutions have found interesting applications,
as for example in the problem of cellular automata (for a review
see Ref. \cite{Rujan}). Moreover, they also serve to built a new
kind  of series expansion for lattice spin systems \cite{Mail}.

\begin{figure}
\begin{center}
\includegraphics[width=1.6 in]{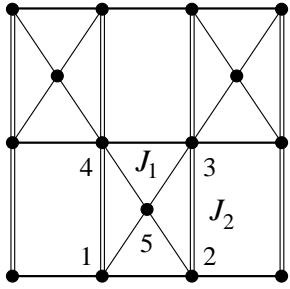}
\end{center}
\caption{ Kagom\'{e} lattice. Interactions between
NN and between NNN, $J_{1}$ and
$J_{2}$, are shown by single and double bonds, respectively.
\label{re-fig7}}
\end{figure}

\begin{figure}
\begin{center}
\includegraphics[width=2.5 in]{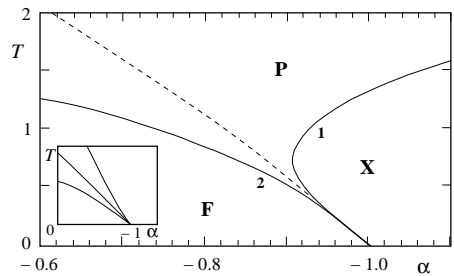}
\end{center}
\caption{ Phase diagram of the Kagom\'{e} lattice
with NNN interaction in the region $J_{1} > 0$ of the space
($\alpha=J_{2}/J_{1}, T$). $T$ is measured in the unit of
$J_{1}/k_{B}$.  Solid lines are critical lines, dashed line is the
disorder line. P, F and X stand for paramagnetic, ferromagnetic
and partially disordered phases, respectively.  The inset shows
schematically enlarged region of the endpoint. \label{re-fig20}}
\end{figure}

Another example is the Kagom\'e lattice shown in Fig. \ref{re-fig7}.  The phase diagram near the phase border is shown in Fig. \ref{re-fig20} with the Stephenson disorder line.  Other very rich phase diagrams are found in Refs. \cite{Diep1987,Diep1991a,Diep1991b,Diep1992} for honeycomb lattice, generalized Kagom\'e lattice and dilute centered square lattice. In some cases, up to five successive phase transitions and several disorder lines are found for a single set of interaction parameters.


\section{Magnetic thin films near the phase frontiers}\label{SM}

The presence of a surface perturbs the bulk properties of a crystal. The perturbation becomes important when the ratio ``number of surface atoms to number of bulk atoms" becomes significative.  Among numerous surface effects, we will outline here  some results on surface magnetization and surface phase transition near  phase frontiers.  The first example which illustrates very well the necessary combination between theory and simulation is the case of a FCC thin film with a (001) frustrated surface \cite{Ngo2007}.
The Hamiltonian is given by
\begin{equation}
\mathcal H=-\sum_{\left<i,j\right>}J_{i,j}\mathbf S_i\cdot\mathbf
S_j -\sum_{<i,j>} I_{i,j}S_i^z S_j^z  \label{eqn:hamil1}
\end{equation}
where $\mathbf S_i$ is the Heisenberg spin at the lattice site
$i$, $\sum_{\left<i,j\right>}$ indicates the sum over the nearest neighbor spin
pairs  $\mathbf S_i$ and $\mathbf S_j$.  The last term, which will
be taken to be very small,  is needed to ensure that there is a phase transition at a finite
temperature for the film with a
finite thickness  when all exchange interactions $J_{i,j}$
are short-ranged. Otherwise, it is known that a
strictly two-dimensional system with an isotropic non-Ising spin
model (XY or Heisenberg model) does not have a long-range ordering
at finite temperatures \cite{Mermin}.  Between the surface spins we take  $J_{i,j}=J_s$, and  between all other spins $J_{i,j}=J=1>0$ (ferromagnetic).  The GS configuration depends on $\eta=J_s/J$. When $\eta$ is smaller than a critical value, the GS becomes non linear. 
We show  $\cos \alpha$ and $\cos \beta $ in Fig. \ref{fig:gscos}
as functions of $J_s$ where $\alpha$ and $\beta$ are the angles between spins defined in the caption. The critical value $J_s^c$ where the collinear configuration becomes non collinear is found between
-0.18 and -0.19.  This value can be calculated analytically (see \cite{Ngo2007}).
\begin{figure}
\begin{center}
\includegraphics[width=2.2 in]{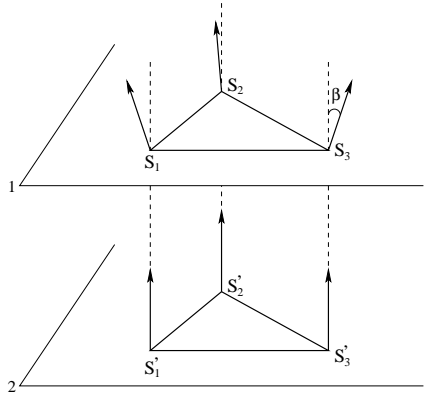}
\end{center}\caption{Non
collinear surface spin configuration. Angles between spins on
layer $1$ are all equal (noted by $\alpha$), while angles between
vertical spins are $\beta$. \label{fig:gsstruct}}
\end{figure}
\begin{figure}
\begin{center}
\includegraphics[width=2.2 in]{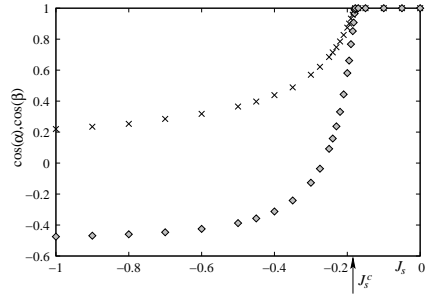}
\end{center}
\caption{$\cos
\alpha$ (diamonds) and $\cos \beta$ (crosses) as functions of
$J_s$. Critical value of $J_s^c$ is shown by the arrow.\label{fig:gscos}}
\end{figure}

\begin{figure}
\begin{center}
\includegraphics[width=2.2 in]{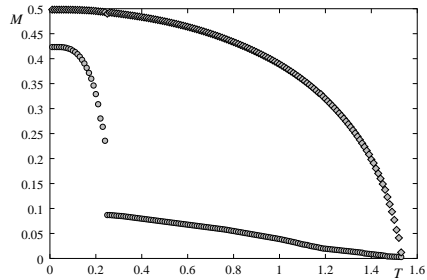}
\end{center}
\caption{First
two layer-magnetizations obtained by the Green's function technique
vs. $T$ for $J_{s} = -0.5$ with $I=-I_s=0.1$. The surface-layer
magnetization (lower curve) is much smaller than the second-layer
one. \label{fig:HGn05Ms}}
\end{figure}

\begin{figure}
\begin{center}
\includegraphics[width=2.2 in]{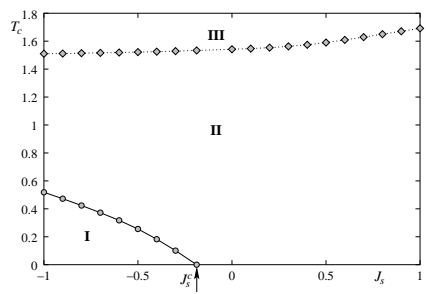}
\end{center}
\caption{Phase
diagram in the space ($J_{s},T$) for the quantum Heisenberg model
in a 4-layer film with $I=|I_s|=0.1$. Phase I: all spins are ordered, phase II: for $J_s<J_s^c$ surface spins are disordered but bulk spins are ordered, for $J_s>J_s^c$ all spins are ordered, phase III: paramagnetic phase. \label{fig:HGDG}}
\end{figure}

Using the theory of Green's function \cite{Diep1979}, we have calculated
the layer magnetizations shown in Fig. \ref{fig:HGn05Ms} (see details in \cite{Ngo2007}).
The phase diagram near the phase boundary $J_s^c$ is shown in Fig. \ref{fig:HGDG}. We notice that the surface
transition occurs only below  $J_s^c$.

MC simulations give very similar results.  Note however that at $T=0$  quantum fluctuations causes
a very strong ``zero-point spin contraction" for surface spins: as a consequence, surface spins do not have
the full length $1/2$ as seen in Fig. \ref{fig:HGn05Ms}, unlike the classical spins used in MC simulations.
\begin{figure}
\begin{center}
\includegraphics[width=2.2 in]{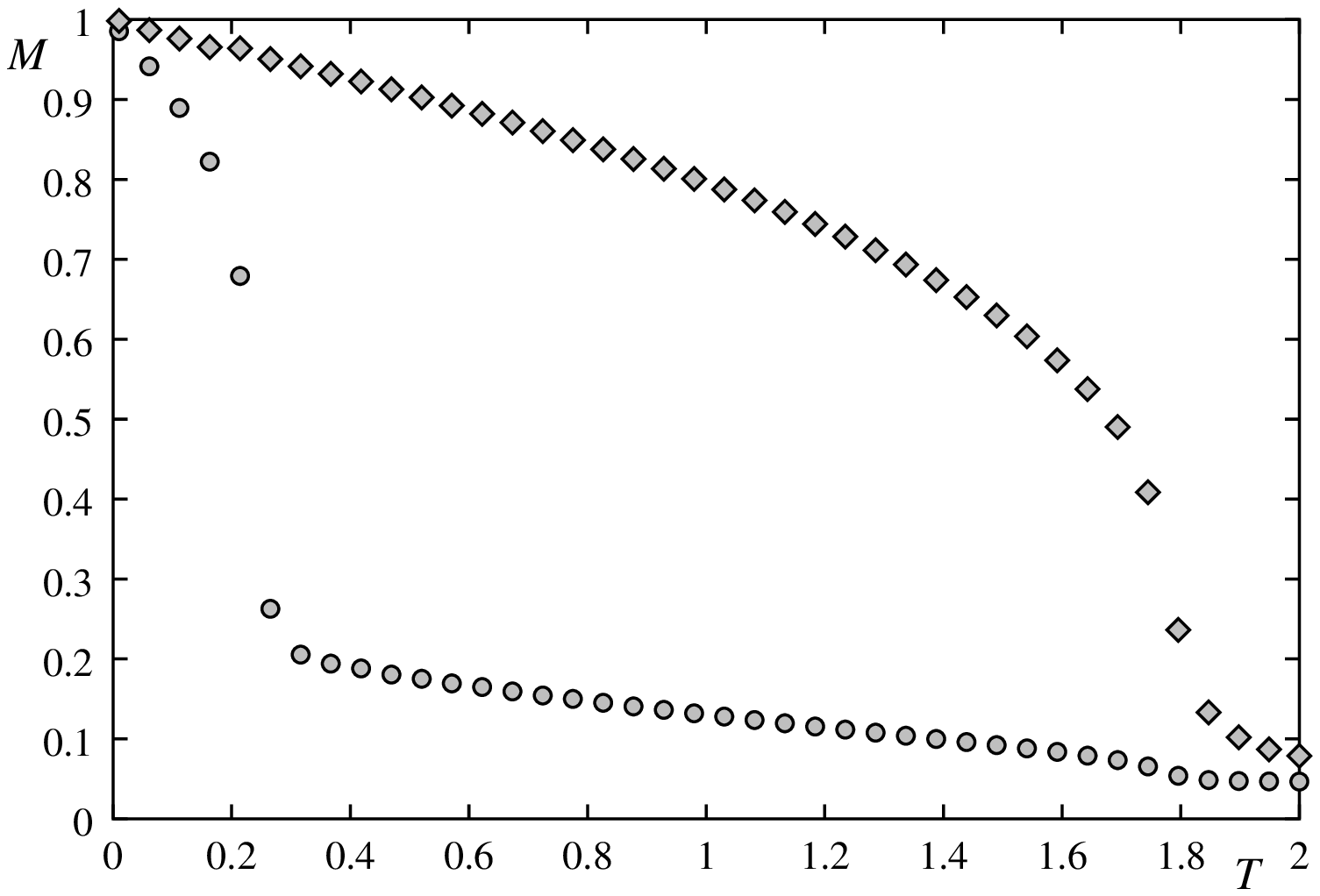}
\end{center}
\caption{Monte Carlo results: magnetizations of layer 1 (circles) and layer 2
(diamonds) versus temperature $T$ in unit of $J/k_B$ for
$J_s=-0.5$ with $I=-I_s=0.1$.} \label{fig:HSn05Ms}
\end{figure}
\begin{figure}
\begin{center}
\includegraphics[width=2.2 in]{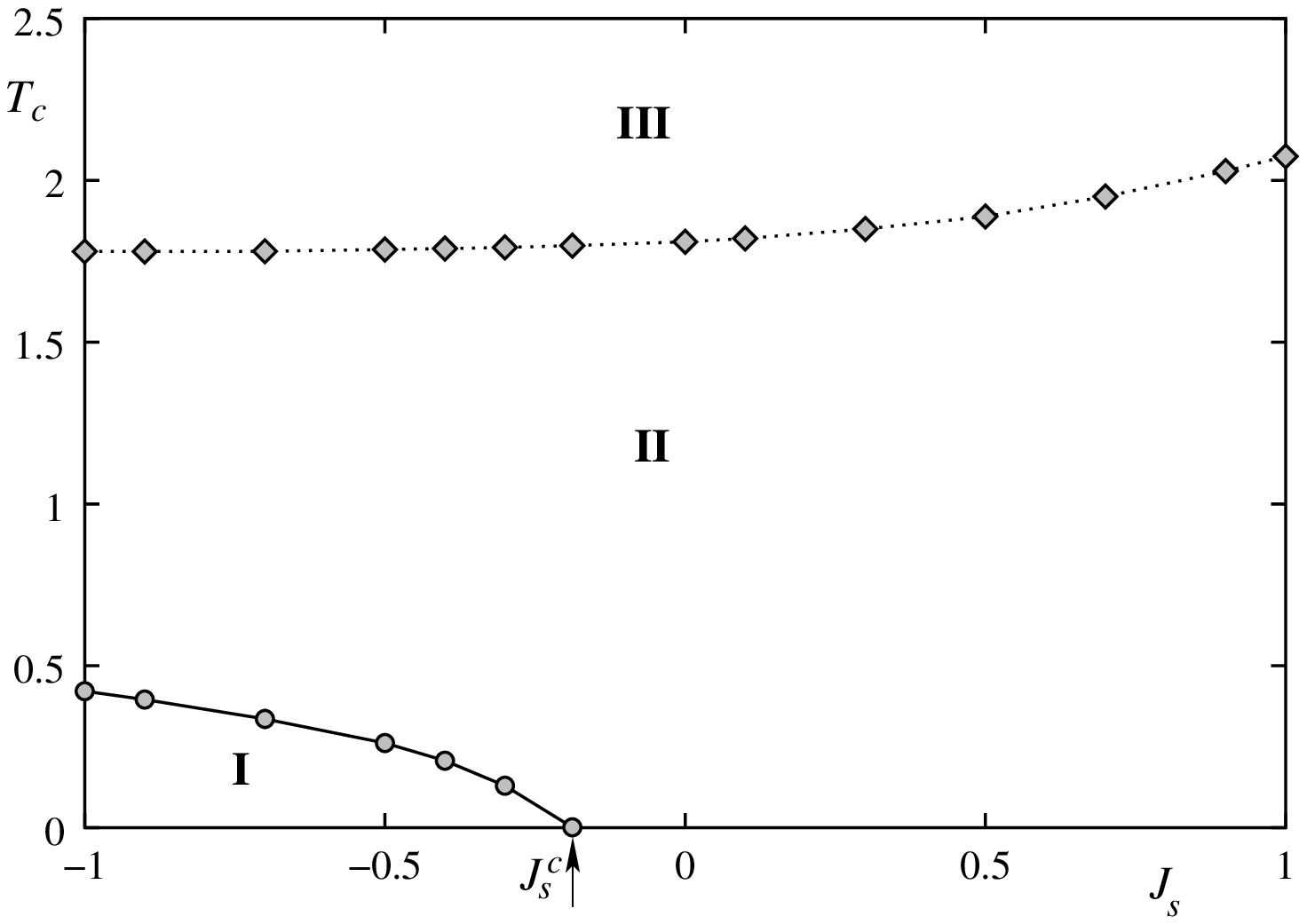}
\end{center}
\caption{Phase
diagram in the space ($J_s,T$) for the classical Heisenberg model
in a 4-layer film with $I=|I_s|=0.1$. Phases I to III have the same
meanings as those in Fig. \ref{fig:HGDG} .} \label{fig:HSDG}
\end{figure}

\begin{figure}
\begin{center}
\includegraphics[width=2.1 in]{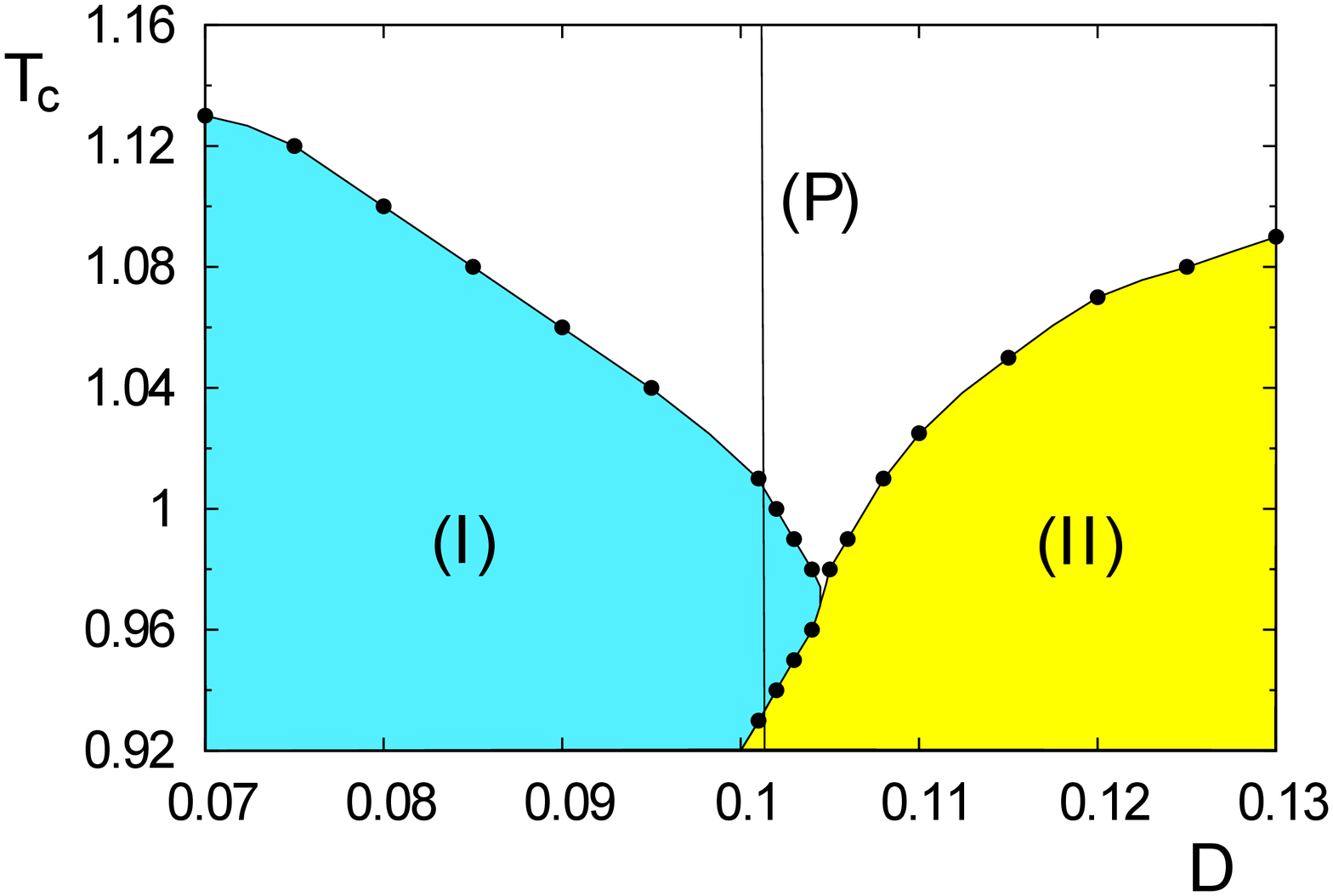}
\includegraphics[width=2.1 in]{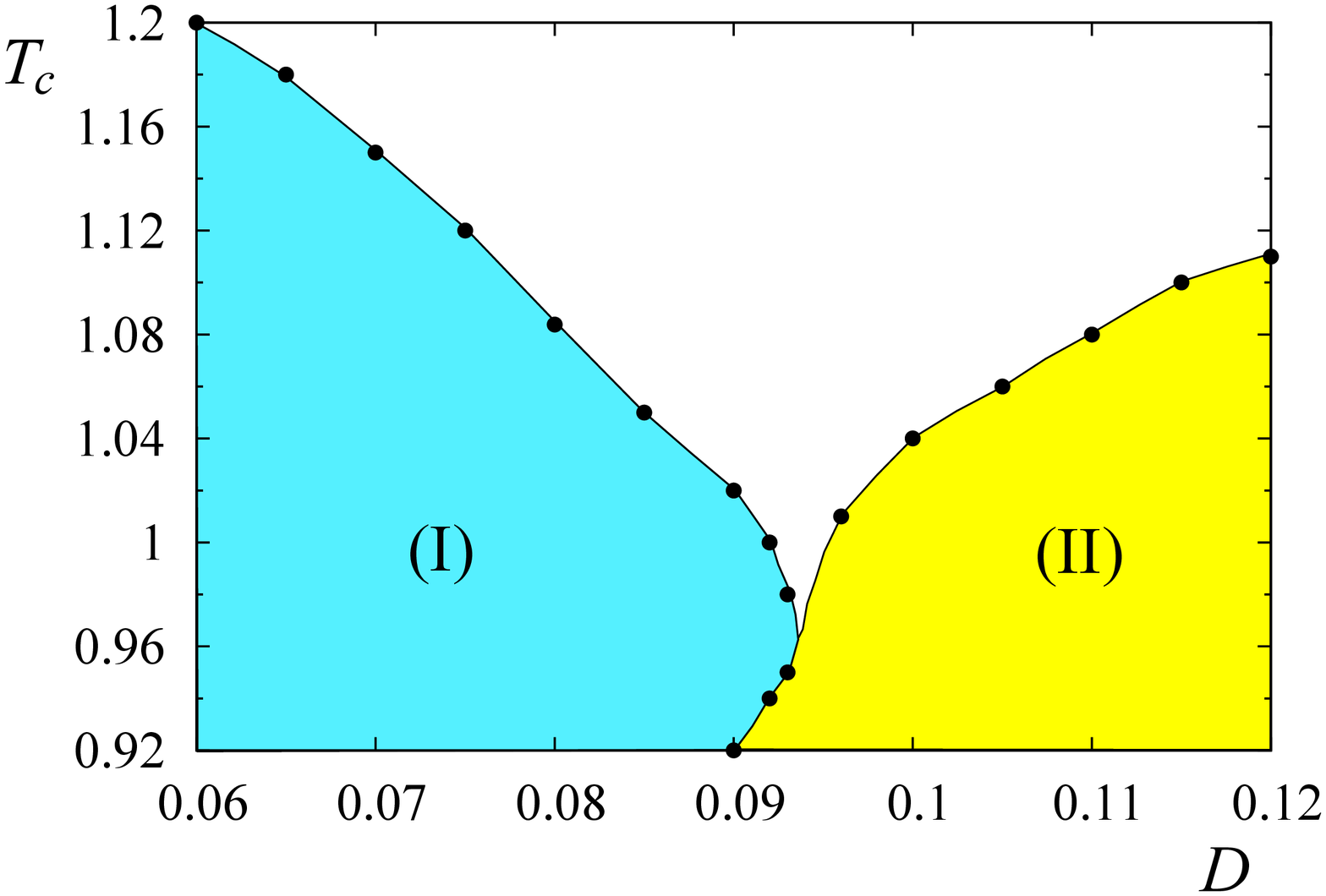}
\end{center}
\caption{(Color online) Phase diagram in 2D: Transition temperature $T_C$ versus $D$, with $A=0.5$, $J=1$ for two dipolar cutoff distances: $r_c=\sqrt{6}$ (left) and $r_c=4$ (right). Phase (I) is the perpendicular spin configuration, phase (II) the in-plane spin configuration and phase (P) the paramagnetic phase. } \label{PD2D}
\end{figure}

The next example is a 2D layer with a dipolar interaction between the 3-state Potts model \cite{Hoang2013}. The competition is between the dipolar term of magnitude $D$, which favors the in-plane spin configuration, and the perpendicular anisotropy $A$ which favors the perpendicular configuration. There is a critical value of $D/A$ above (below) which the configuration is planar (perpendicular). Near this frontier, we found by MC simulations an interesting phenomenon which is called the re-orientation phase transition shown in Fig. \ref{PD2D}: for example, if we follow the vertical line on the left figure, we see that at low $T$ the system is in the planar configuration (phase II), 
 it crosses the phase separation line with increasing $T$ to go to the perpendicular phase (phase I)
before going out to the paramagnetic phase (phase P) at a higher $T$.  Such a re-orientation 
at a finite $T$ is spectacular: we have shown that the transition between phases I and II is of first order \cite{Hoang2013}. This is not similar to the reentrant region shown
in Figs. \ref{re-fig18} and \ref{re-fig20} where all transition lines are of second order and a very narrow paramagnetic phase separates
the two ordered phases (AF and F, or F and X).  It is interesting to note that to allow a transition 
between two ``incompatible" symmetries (the one is not a subgroup of the other), 
there are two ways: i) a single first-order transition with a latent heat (as the case shown in Fig. \ref{PD2D}), ii) two second-order
transitions separated by a narrow region of a reentrant paramagnetic phase with often a disorder 
line separating two zones of different pre-ordering fluctuations (as the case shown in Fig. \ref{re-fig20}). 
Note that the re-orientation transition has also been observed for the Heisenberg spin model \cite{Santa2000}. 

\begin{figure}
\begin{center}
\includegraphics[width=1.8 in,angle=-90]{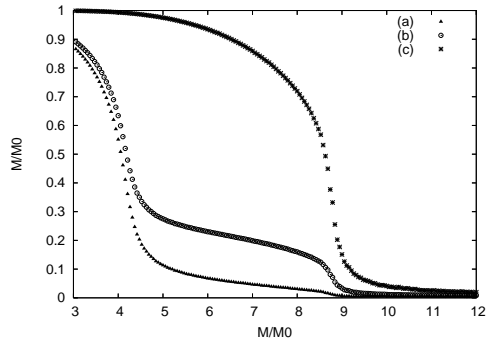}
\end{center}
\caption{Magnetization  versus $T$ in the case where
the system is made of three films (see text).  Black
triangles: magnetization of the surface films, stars:
magnetization of the middle film, void circles: total
magnetization. }\label{surf1}
\end{figure}

\begin{figure}
\begin{center}
\includegraphics[width=2.2 in]{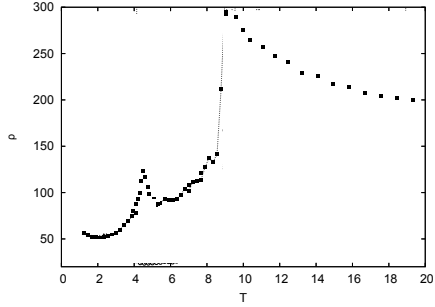}
\end{center}
\caption{ Resistivity $\rho$ in arbitrary unit versus  $T$ of the
system described in the previous figure's caption.}\label{surf2}
\end{figure}

Let us give now an example where the surface disordering affects strongly the spin resistivity in a magnetic thin film.  The spin resistivity has been studied both theoretically and experimentally for more than 50 years. The reader is referred to Refs. \cite{Akabli2008,Magnin2011,Magnin2012} for references.  Theoretically, the coupling between an itinerant spin with lattice spins affects strongly the spin resistivity $\rho$.  The spin-spin correlation has been shown to be the main mechanism which governs $\rho$.
When there is a magnetic phase transition, the spin resistivity undergoes an anomaly. In magnetic thin films, when there is a surface phase transition at a temperature $T_s$ different
from that of the bulk one ($T_c$), we expect two peaks of $\rho$ one at $T_s$ and the other at $T_c$.
We show here an example of a thin film, of FCC structure with Ising spins, composed of three sub-films:  the middle film of
4 atomic layers between two surface films of 5 layers. The lattice sites are occupied by Ising spins interacting with   each other via NN ferromagnetic interaction.  Let us suppose the interaction between spins in the outside films be $J_s$ and that in the middle film be $J$. The inter-film interaction is $J$.
In order two enhance surface effects we suppose  $J_s<<J$.   We show
in Figs. \ref{surf1} and \ref{surf2} the layer magnetization and the spin resistivity for  $J_s=0.2 J$. We observe that the surface films undergo a phase
transition at $ T_s \simeq 4$ far below the transition temperature of the middle film $T_c \simeq 8.8$.  As stated above,
a phase transition induces an anomaly in the spin resistivity: the two phase transitions observed
in Fig. \ref{surf1} give rise to two peaks of $\rho$ shown in Fig. \ref{surf2}.  The surface peak of $\rho$   has been also seen in a  frustrated film \cite{Magnin2011}.


\begin{figure}
\begin{center}
\includegraphics[width=2.2 in]{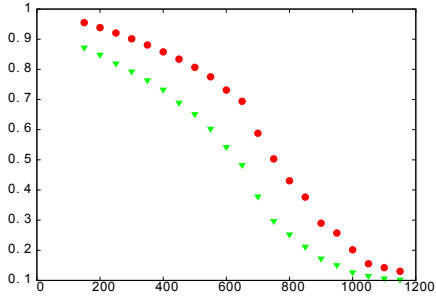}
\end{center}
\caption{\label{structure_factor}(Color online) Structure factor $S_{\vec K}$ (green triangles) and  $O_6$ order parameter (red circles) of the first layer versus temperature for the EAM potential.}
\end{figure}

\begin{figure}
\begin{center}
\includegraphics[width=2.3 in]{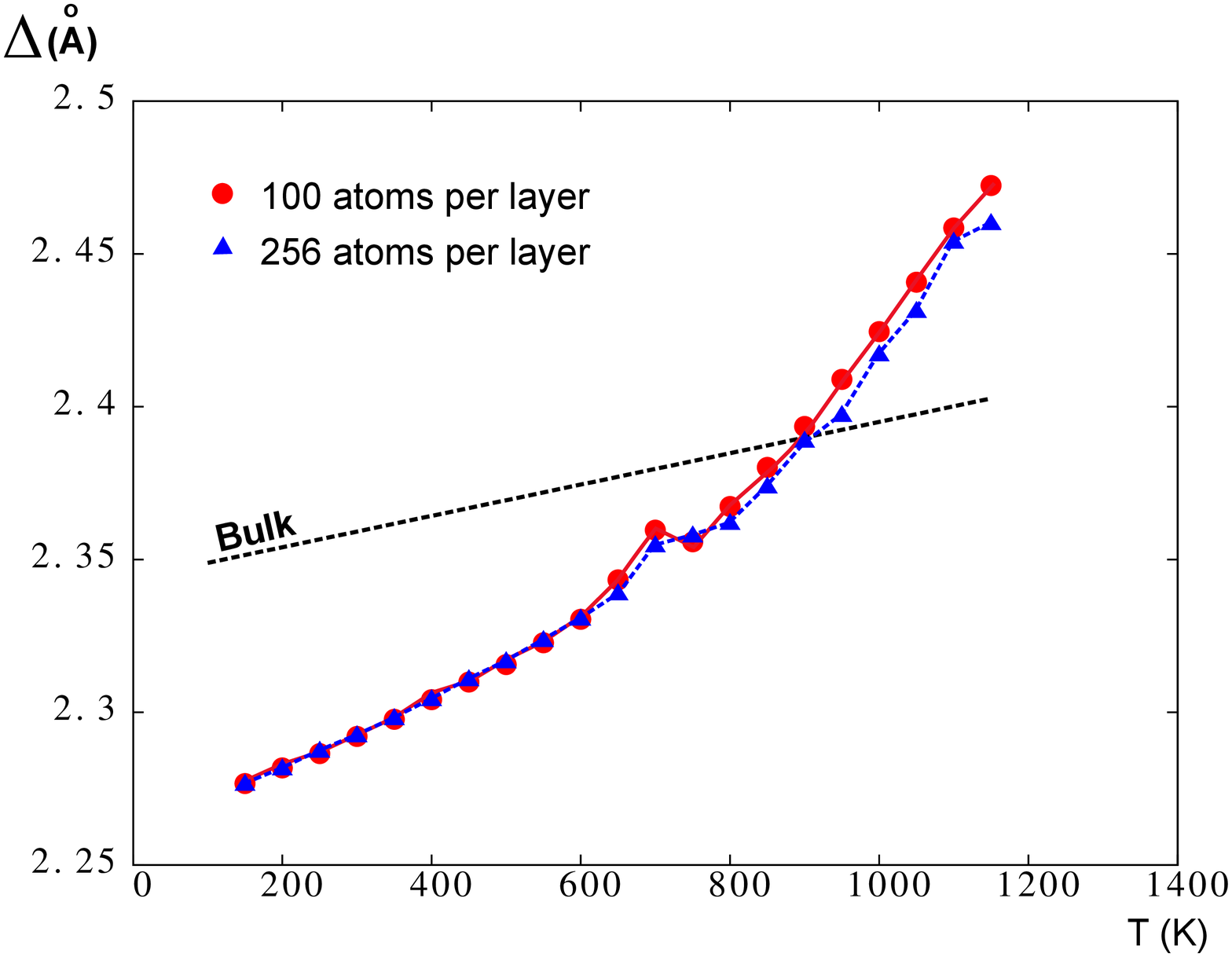}
\end{center}
\caption{\label{surfacesize}(Color online) Surface relaxation for two surface sizes 100 and 256 atoms for the EAM potential. }
\end{figure}

The last example is the surface relaxation and the surface melting of  a semi-infinite Ag crystal with a (111) surface \cite{Bocchetti2013b}. The border here, contrary to the previous examples, is a physical border (not a phase border). The competition exists also for this case because surface and bulk atoms have different environments and the multi-body interactions compete with the two-body terms in the potential from the Embedded-Atom-Method (EAM) \cite{Zhou2001} we have used.  In order to see the surface melting,  we compute the structure factor $S_{\vec K}$  as follows:

\begin{equation}\label{SFdef}
S_{\vec K} =\frac{1}{N_l}\left < \left |\sum_{j=1}^{N_l}e^{i\overrightarrow{K}.\overrightarrow{d_j}}\right |\right >
\end{equation}
where $\overrightarrow{d_j}$ is the position vector of an atom in the layer, $N_l$  the number of atoms in a layer and $\overrightarrow{K}$  the reciprocal lattice vector which has the following coordinate (in reduced units):
$2\pi\left(1;-\sqrt{3};0\right)$.  The angular brackets $<...>$ indicate thermal average taken over MC run time.
The above  ``order parameter", which allows us to monitor the long-range surface order,
is plotted for the surface layer in Fig. \ref{structure_factor}.
 As we can see, the long-range order is lost  at $\simeq 700$ K.  Note that the bulk Ag melts at $\simeq 1235$ K.
 In order to investigate in more details the surface melting, we have also computed the $O_6$ order parameter which describes the short-range hexatic orientational
order of the surface :

\begin{equation}
O_6 =\frac{\left|\sum_{jk}W_{jk}e^{i6\Theta_{jk}}\right|}{\sum_{jk}W_{jk}}
\end{equation}
with
\begin{equation}
W_{jk} = e^{- \frac{\left(z_j-z_k\right)^2}{2\delta^2}}
\end{equation}
where the sum runs over the NN pairs and $\Theta_{jk}$ is the angle which the $j-k$ bond, when projected on the $xy$ plane, forms with the $x$ axis. The $\delta$ parameter is taken as one-half the average inter-layer spacing.
The weighting function, $W_{jk}$, allows us to differentiate the ``non coplanar" and the ``coplanar" neighbors. With a coplanar neighbor, the weighting function takes a maximum value.
We have to calculate the spatial average of $O_6$ taken over all atoms of the surface layer and then calculate its thermal
average over MC run time. We plot the averaged $O_6$ parameter   
versus temperature in Fig. \ref{structure_factor}.  The short-range order is also lost at 700 K.
We have calculated the distance $\Delta$ between the topmost layer and the second layer. There is a contraction of this distance with respect to the distance between two layers in the bulk as seen in
Fig. \ref{surfacesize}.  Only at about 900 K, far after the surface melting, that surface 
atoms are ``desorbed" from the crystal.




\section{Concluding remarks}\label{Concl}
We have shown in this paper a number of phenomena occurring at frontiers between phases of different symmetries.  The phase diagram near the phase frontiers is often very rich with reentrance, disorder lines and multiple phase transitions.  One has seen that in most of the cases treated so far we need a combination of theory and simulation to better understand complicated physical effects resulting from competing forces around the frontiers.  A frontier is determined as a compromise between these forces. As a consequence, frontiers  do not have a high stability: we have seen that when a small external perturbation, such as the temperature, is introduced, the phase border moves in favor of one of the neighboring phases according to some criteria such as the entropy.  Therefore,  many interesting effects manifest themselves  around frontiers. The physics near phase frontiers is far from being well understood in various situations.

\ack
H T Diep thanks the IOP-Hanoi for a financial support. V T Ngo is indebted to Vietnam National Foundation for Science and Technology Development  (Nafosted)  for the Grant No. 103.02-2011.55.


\section*{References}
\begin{thebibliography}{9}
\bibitem{DiepFSS2013} Diep H T (ed) 2013 {\it Frustrated Spin Systems} 2nd edition (Singapore: World Scientific)
\bibitem{Diep1987}Azaria P, Diep H T and Giacomini H 1987 {\it Phys. Rev. Lett.} {\bf 59} 1629
\bibitem{Diep1991a} Diep H T, Debauche M and Giacomini H 1991{\it Phys. Rev.} B {\bf 43} 8759
\bibitem{Diep1991b} Debauche M, Diep H T, Azaria P and Giacomini H 1991 {\it Phys. Rev.} B {\bf 44} 2369
\bibitem{Diep1992}  Debauche M and Diep H T 1992 {\it Phys. Rev. } B {\bf
46} 8214 ; Diep H T,  Debauche M and Giacomini H 1992 {\it
J. of Mag. and Mag. Mater.} {\bf 104} 184
\bibitem{Diep1979} Diep-The-Hung, Levy J C S  and Nagai O 1979  {\it Phys. Stat. Sol.} (b)  {\bf 93} 351
\bibitem{Ngo2007} Ngo V T and Diep H T 2007 {\it Phys. Rev. } B {\bf 75} 035412
\bibitem{Hoang2013} Hoang D-T, Kasperski M, Puszkarski H, and Diep H T 2013 {\it J. Phys.: Cond. Matter} {\bf 25} 056006
\bibitem{Akabli2008} Akabli K and Diep H T 2008 {\it Phys. Rev.} B {\bf 77} 165433
\bibitem{Magnin2011} Magnin Y, Akabli K and Diep H T 2011 {\it Phys. Rev.} B {\bf 83} 144406
\bibitem{Bocchetti2013b} Bocchetti V and Diep H T 2013 {\it Surf. Sci. }  {\bf 614} 46

\bibitem{Vaks} Vaks V,  Larkin A and Ovchinnikov Y 1966 {\it Sov. Phys. JEPT}
{\bf 22}  820

\bibitem{Aza89a}  Azaria P,  Diep H T and  Giacomini H 1989 {\it Phys. Rev. }
B {\bf 39} 740

\bibitem{Blan}  Blankschtein D,  Ma M and Berker A N 1984 {\it Phys. Rev. } B
{\bf 30}  1362
\bibitem{Diep85b} Diep H T, Lallemand P and Nagai O 1985 {\it J. Phys. } C
{\bf 18} 1067

\bibitem{Quartu1997} Quartu R and Diep H T 1997 {\it Phys. Rev.} B {\bf 55} 2975

\bibitem{Santa1997}  Santamaria C and Diep H T 1997 {\it J. Appl. Phys.} {\bf 81} 5276

\bibitem{Ste} Stephenson J 1970 {\it J. Math. Phys.} {\bf 11}  420 ; ibid 1970
{\it Can. J. Phys.} {\bf 48}  2118 ; ibid. 1970 {\it Phys. Rev. } B
{\bf 1} 4405

\bibitem{Mail} Maillard J M 1986 {\it Second Conference on Statistical Mechanics},
California Davies, unpublished.

\bibitem{Giaco86} Giacomini H 1986 {\it J. Phys. A} {\bf 19}  L335
\bibitem{Rujan}Rujan P 1987 {\it J. Stat. Phys.} {\bf 49} 139

\bibitem{Mermin} Mermin N D and Wagner H 1996, {\it Phys. Rev. Lett.} {\bf 17} 1133


\bibitem{Santa2000}  Santamaria C and Diep H T 2000 {\it J. Mag. Mag. Mater.} {\bf 212} 23

\bibitem{Magnin2012} Magnin Y and Diep H T 2012 {\it Phys. Rev.  } B {\bf 85} 184413

\bibitem{Zhou2001} Zhou X W,  Wadley H N G, Johnson R A, Larson D J, Tabat N, Cerezo A,  Petford-Long A K, Smith G D W, Clifton P H, Martens R L and Kelly T F 2001 {\it Acta Materialia } {\bf 49} 4005
\end{thebibliography}
\end{document}